\documentclass[namedreferences]{solarphysics}

\usepackage[T1]{fontenc}
\usepackage[utf8]{inputenc}

\usepackage[optionalrh,hyperref,solaromanenum,linksfromyear]{spr-sola-addons}
\usepackage{graphicx}
\usepackage{color}
\usepackage{breakurl}

\hypersetup{
    colorlinks=true,
    linkcolor=blue,
    citecolor=blue,
    filecolor=black,
    urlcolor=blue,
}

\begin{document}
\begin{article}
\begin{opening}

\title{Active Region Contributions to the Solar Wind Over Multiple Solar Cycles}

\author[addressref={1},corref,email={d.stansby@ucl.ac.uk}]{\inits{D }\fnm{David }\lnm{Stansby}\orcid{0000-0002-1365-1908}}
\author[addressref={1}]{\inits{L M}\fnm{Lucie M. }\lnm{Green}\orcid{0000-0002-0053-4876}}
\author[addressref={1,2,3}]{\inits{L}\fnm{Lidia }\lnm{van Driel-Gesztelyi}\orcid{0000-0002-2943-5978}}
\author[addressref={4}]{\inits{T S}\fnm{Timothy S. }\lnm{Horbury}\orcid{0000-0002-7572-4690}}

\address[id={1}]{Mullard Space Science Laboratory, University College London, Holmbury St. Mary, Surrey, RH5 6NT, UK}
\address[id={2}]{LESIA, Observatoire de Paris, Universit\'e PSL, CNRS, Sorbonne Universit\'e, Universit\'e Paris Diderot, Sorbonne Paris Cit\'e, 92195 Meudon, France}
\address[id={3}]{Konkoly Observatory, Research Centre for Astronomy and Earth Sciences, Konkoly Thege u.15-17., H-1121, Budapest, Hungary}
\address[id={4}]{Imperial College London, South Kensington Campus, London, SW7 2AZ, UK}

\runningauthor{D.~Stansby et al.}
\runningtitle{Active Region Solar Wind Sources}

\begin{abstract}
Both coronal holes and active regions are source regions of the solar wind. The distribution of these coronal structures across both space and time is well known, but it is unclear how much each source contributes to the solar wind. In this study we use photospheric magnetic field maps observed over the past four solar cycles to estimate what fraction of magnetic open solar flux is rooted in active regions, a proxy for the fraction of all solar wind originating in active regions. We find that the fractional contribution of active regions to the solar wind varies between 30\% to 80\% at any one time during solar maximum and is negligible at solar minimum, showing a strong correlation with sunspot number. While active regions are typically confined to latitudes $\pm$30$^{\circ}$ in the corona, the solar wind they produce can reach latitudes up to $\pm$60$^{\circ}$. Their fractional contribution to the solar wind also correlates with coronal mass ejection rate, and is highly variable, changing by $\pm$20\% on monthly timescales within individual solar maxima. We speculate that these variations could be driven by coronal mass ejections causing reconfigurations of coronal magnetic field on sub-monthly timescales.
\end{abstract}

\end{opening}

\section{Introduction}
The solar wind is a flow of hot tenuous plasma, driven by the large pressure difference between the Sun's corona and the interplanetary medium. Not all areas of the corona escape to form the solar wind: in some areas plasma is confined on closed field lines, whereas in others the plasma accelerates until it becomes super-sonic and super-Alv\'enic, carrying magnetic flux out into interplanetary space to form the solar wind.

The global properties of the solar wind vary with, and are ultimately controlled by, the Sun's 11 year activity cycle \citep[e.g.][]{McComas2013}. At the beginning of a cycle, during solar minima, the corona is dominated by two polar coronal holes at high latitudes. These host open magnetic field lines, while at low latitudes closed loops dominate. This relatively simple configuration is disrupted by strong concentrations of magnetic flux emerging through the photosphere at low latitudes, forming new active regions \citep[ARs,][]{vanDriel-Gesztelyi2015, Cheung2017}. As magnetic flux emerges through the photosphere it starts out closed, but as the field strength increases the closed loops can reconnect with adjacent open field lines \citep{Driel-Gesztelyi2014, Ma2014, Kong2018},  redistributing regions of existing open flux \citep{Sheeley1989, Baker2007} and in the process opening up previously closed flux \citep{Wang2003a, Attrill2006}. This allows plasma originating in active regions to flow out from the corona and form part of the solar wind.

As time passes, photospheric footpoints of open field lines are subject to diffusion processes, causing the initially high concentrations of magnetic flux to spread out and become weaker \citep{Leighton1964}. As the magnetic field strength weakens these areas remain open and turn into low latitude equatorial coronal holes \citep{Wang2010a, Petrie2013a, Wang2017a, Golubeva2017}. As the solar cycle progresses further the dominant polarity of each polar coronal hole is eroded, and new polar coronal holes of opposite polarity form. Eventually the rate of flux emergence at low latitudes decreases, leaving these two new polar coronal holes and marking the beginning of the subsequent cycle.

The idea that active regions could be source regions for solar wind was initially developed via global coronal models. These models predict where open flux is rooted, and in several individual cases predicted significant amounts of open flux rooted in active regions \citep{Neugebauer2002, Wang2003}. This was subsequently backed up by evidence showing persistent upflows of coronal plasma on the edges of active regions \citep[see][for a review]{Tian2021}. Estimates of heavy ion composition within active regions match measurements in the solar wind, providing further evidence for active region contributions to the solar wind \citep{Macneil2019, Stansby2020d}.

While the spatial and temporal distribution of active regions has been known for a long time, how this translates into contributions to the solar wind is less well known. \cite{Schrijver2003} used global coronal potential field source surface models to show an increase in the fraction of open flux rooted in active regions during the rising phase of Cycle 23, peaking at $\sim$30\% during solar maximum (their Figure 11). Using a slightly different method, and considering only solar wind in the ecliptic plane \cite{Fu2015} concluded that ARs were the major contributor to solar wind measured at Earth during the maximum of Cycle 23, with the contribution declining from $\sim$60\% to $\sim$10\% during the declining phase of that cycle.

It is not clear whether these results are dependent on magnetogram data source (both \cite{Schrijver2003} and \cite{Fu2015} only used a single source), and how the variation extends over multiple solar cycles. In this paper we perform such an analysis of active region solar wind sources, using four different magnetogram sources, allowing us to span the last four solar cycles, and check that different data sources agree when they overlap.

In Section \ref{sec:methods} the data and methods used to distinguish between active region and coronal hole sources are explained. Results are presented in Section \ref{sec:results}, showing that the fractional contribution of active regions to the solar wind can be up to 80\% during solar maximum, showing a strong correlation with sunspot number. These results are placed in context and discussed in Section \ref{sec:discussion}, with conclusions given in Section \ref{sec:conclusions}.

\section{Methods}
\label{sec:methods}
Because both the heliospheric magnetic flux and solar wind mass flux are approximately independent of latitude and longitude in the heliosphere \citep{Lockwood2009, Wang2010, McComas2013}, measuring the fraction of open magnetic flux rooted within different coronal source regions is a convenient proxy for the amount of solar wind originating in different source regions. In the following subsections we give an overview of how the photospheric footpoints of open flux were computed (Section \ref{sec:pfss}) and how coronal hole and active region areas were distinguished (Section \ref{sec:threshold}).

\subsection{PFSS Modelling}
\label{sec:pfss}
To estimate where open flux is rooted in the photosphere, potential field source surface \citep[PFSS,][]{Altschuler1969, Schatten1969} modelling was carried out using the \texttt{pfsspy} software package \citep{Stansby2020e}. Several observatories provide synoptic magnetic field maps, using different observing equipment, techniques, and data processing pipelines \citep[see e.g.][Table 1 for a list]{Riley2014}. In this study synoptic maps from the Kitt Peak Vacuum Telescope \citep[KPVT,][]{Livingston1976}, Michelson Doppler Interferometer \citep[MDI,][]{Scherrer1995}, Synoptic Optical Long-term Investigations of the Sun \citep[SOLIS,][]{Keller1998}, and Global Oscillations Network Group \citep[GONG,][]{Harvey1996} were used. Between them these data sources provide synoptic maps of the radial component of magnetic field in the photosphere, spanning more than four solar cycles at a rate of one map per Carrington rotation. KPVT, SOLIS, and GONG maps are provided at a native resolution of 360 $\times$ 180 (longitude $\times$ $\sin$ latitude), and MDI maps were re-binned from 3600 $\times$ 1080 to this lower resolution. Table \ref{tab:sources} lists the data products along with their temporal coverage and links to the data.
\begin{table}
\caption{Details of the data sources used in this study. Text in the ``Data'' column are clickable http or ftp links.
}
\label{tab:sources}
\begin{tabular}{cccc} 
\hline
Observatory	& Date range	& AR threshold (G)		& Data								 	\\
\hline		
KPVT		& 1975 Feb.~19 -- 2003 Aug.~29	& 35					& \href{ftp://solis.nso.edu/kpvt/synoptic/mag/}{Link} 	 \\
MDI			& 1996 Jun.~28 -- 2010 Nov.~26	& 100				& \href{http://soi.stanford.edu/magnetic/synoptic/carrot/M_Corr/}{Link} \\
SOLIS		& 2003 Aug.~16 -- 2017 Oct.~23	& 40					& \href{ftp://solis.nso.edu/integral/kbv7g}{Link}		 \\
GONG		& 2006 Oct.~18 -- 2021 Apr.~14	& 30					& \href{https://gong2.nso.edu/oQR/zqs/}{Link}		\\
\hline
\end{tabular}
\end{table}

PFSS solutions for individual magnetograms were calculated on a 360 $\times$ 180 grid in (longitude $\times$ $\sin$ latitude) and 50 grid points in the radial direction between the solar surface ($R_{\odot}$) and source surface ($R_{ss}$). Magnetic field lines were then traced from an evenly spaced 360 $\times$ 180 grid at $R_{ss}$ down to $R_{\odot}$. For each individual field line this resulted in six quantities: the source surface latitude and longitude from which the field line was traced ($\theta_{ss}, \phi_{ss}$), the solar surface footpoint latitude and longitude ($\theta_{\odot}, \phi_{\odot}$), and the radial magnetic field strength, both on the source and solar surfaces ($B_{r,ss}, B_{r,\odot}$). The resulting dataset is available at \url{https://doi.org/10.5281/zenodo.4783720}, and the code used to generate it at \url{https://doi.org/10.5281/zenodo.4787880}.

\begin{figure} 
	\centerline{\includegraphics[width=\textwidth]{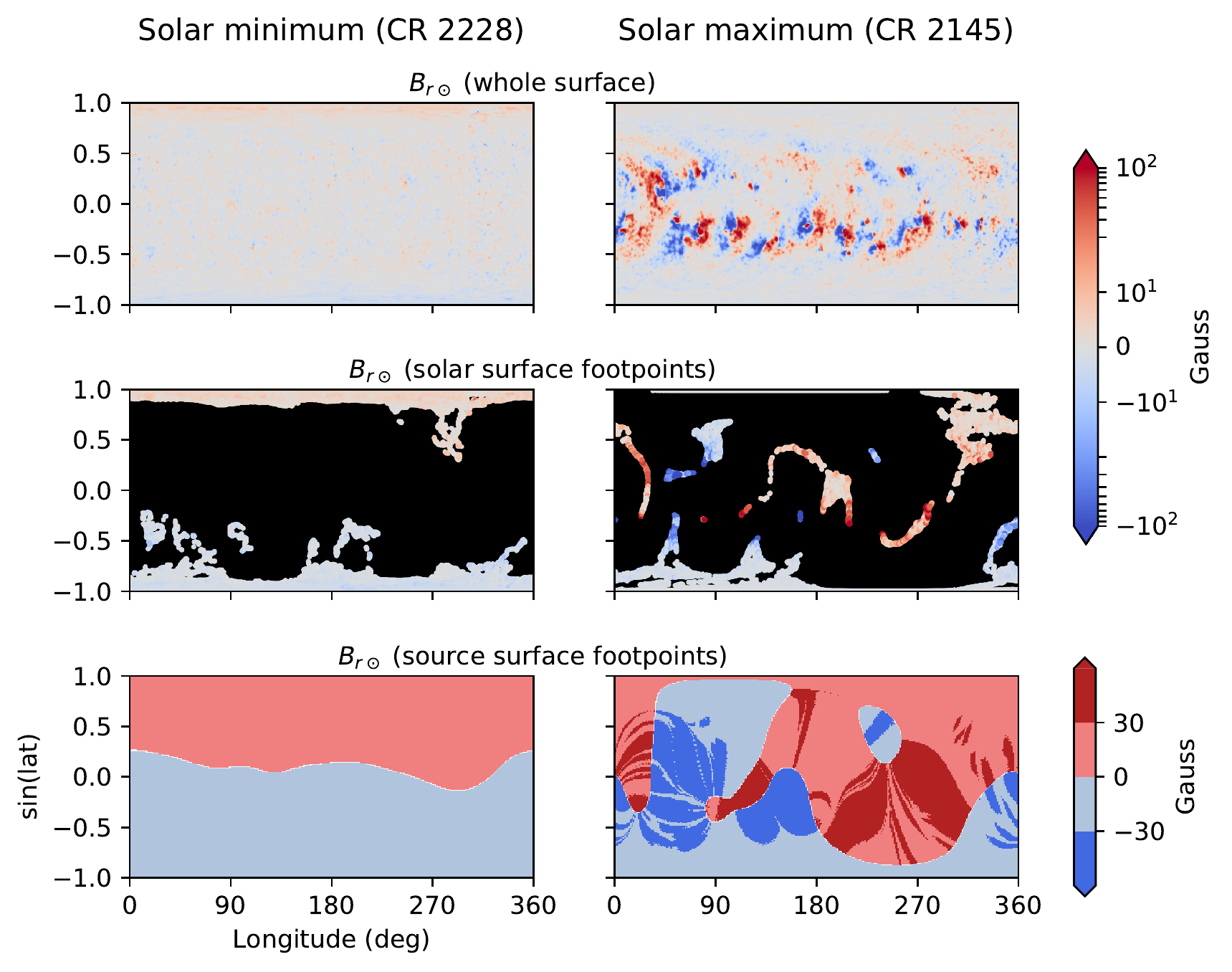}}
	\caption{Photospheric magnetic field maps and open field regions at a solar minimum (left column) and a solar maximum (right column). Top rows show the photospheric magnetic field map. Middle rows show the open field footpoints from field lines traced down from an equal area grid at the source surface. The points are coloured by their photospheric field polarity and strength. The bottom row shows categorisation of source surface magnetic field lines into coronal hole (light colours) and active region (dark colours) sources. Both input magnetograms are from GONG.}
	\label{fig:case study}
\end{figure}

Two examples of this methodology are shown in Figure \ref{fig:case study}, one during solar minimum (left hand panels) and one during solar maximum (right hand panels). The top two panels show the photospheric maps used as input to the PFSS modelling, and the middle two panels show the photospheric footpoints of open field lines, coloured by $B_{r,\odot}$. The bottom panels show the partitioning of open field lines into active region and coronal hole regions; for more details see Section \ref{sec:threshold}.

\subsubsection{Choice of Source Surface Height}
The source surface height ($R_{ss}$) is a parameter of PFSS models that must be chosen to calculate a solution. Various methods can been used to choose an optimal source surface height, where optimal is defined relative to a given observational signature. Examples include matching total unsigned open flux to that measured in-situ in the solar wind \citep[e.g.][]{Lee2011, Arden2014, Virtanen2020}, matching heliospheric current sheet crossings to those measured in-situ \citep[e.g.][]{Hoeksema1983, Badman2020}, or matching the locations of large open field regions with observations of coronal holes in extreme ultra-violet (EUV) images \citep[e.g.][]{Asvestari2019}. `Optimal' source surface heights vary between methods and location within the solar cycle, but $R_{ss}$ almost always lies somewhere in the range $[1.5, 3.0]R_{\odot}$.

To test the robustness of our results against changes in $R_{ss}$, all analysis was run for four source surface heights, $R_{ss} = \{1.5, 2.0, 2.5, 3.0\}R_{\odot}$. The key quantitive result of this paper is not significantly affected by changes in the source surface height (demonstrated later in appendix \ref{app:varying rss}), so for simplicity all results in the main body are shown for $R_{ss} = 2.0 R_{\odot}$. We have inspected a full set of figures from the analysis run at each height to confirm that qualitative conclusions do not change either.

\subsection{Distinguishing Coronal Holes and Active Regions}
\label{sec:threshold}
A key observational difference between coronal holes and active regions is their photospheric magnetic field strength. Coronal holes contain weak fields, compared to active regions with much stronger fields. This makes it possible to set a threshold below which a photospheric footpoint is considered rooted in a coronal hole, and above which it is rooted in an active region.

\begin{figure} 
	\centerline{\includegraphics[width=\textwidth]{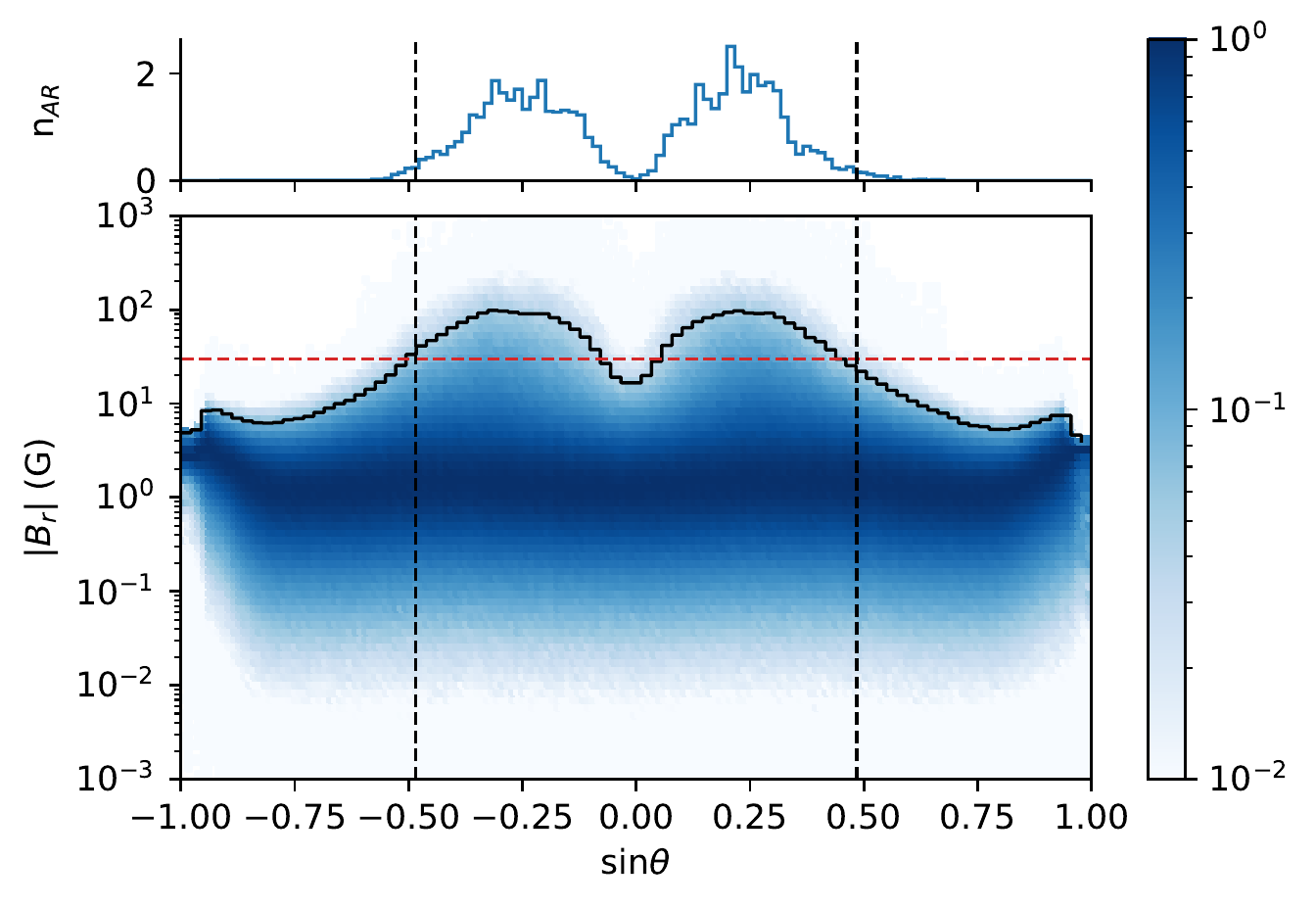}}
	\caption{Distribution of photospheric radial magnetic field strength as a function of latitude from GONG synoptic maps. The data are a complete set of GONG maps from 8 January 2007 -- 25 December 2020 sampled once per Carrington rotation. Solid black line shows the 99th percentile of magnetic field values. The top panel shows the distribution of NOAA active region latitudes for the same time period. Vertical dashed lines denote the 1st and 99th percentile of AR latitudes. The red horizontal line is drawn at 30 G, the threshold chosen to separate coronal hole and active region areas for GONG.}
	\label{fig:b_threshold}
\end{figure}

To aid in choosing such a threshold, Figure \ref{fig:b_threshold} shows the distribution of open footpoint field strengths as a function of latitude for all GONG magnetograms used here. As expected the strongest fields occur at mid-latitudes, where active regions are present. To confirm this, the top panel of Figure \ref{fig:b_threshold} shows the distribution of NOAA active region latitudes during the interval spanned by the magnetic field maps, with the 1st and 99th percentile AR latitudes indicated with vertical dashed black lines.

Outside active region latitudes, it is assumed that all open field footpoints fall within coronal holes. A lower limit on active region magnetic field strength can therefore be set as the maximum field strength outside active region latitudes. Figure \ref{fig:b_threshold} demonstrates how this threshold was chosen for GONG. Tracing the 1st and 99th percentile of AR latitudes down to distributions of magnetic field strength, and then finding the intersection with the 99th percentile or $B_{r,\odot}$ at these latitudes gives a threshold of 30 G, as indicated by the horizontal red line. Because the magnitude of magnetic fields measured by different observatories are systematically different \citep{Riley2014, Virtanen2017a}, this threshold identification process was repeated for each magnetogram source, with the thresholds reported in Table \ref{tab:sources}. These thresholds are all above published estimates for noise levels, which are around 5 G for MDI \citep{Liu2012a}, 5G for KPVT \citep{Wenzler2004}, 1 G for SOLIS \citep{Harvey2003}, and 3 G for GONG \citep{Clark2003}.

While these thresholds may seem low for an active region, they are averaged over a 1 deg$^{2}$ area of the photosphere, washing out the peak values present at smaller scales. In addition, the thresholds used here are similar to thresholds used for assimilating newly observed active regions in the literature, which variously are 15 G \citep{Yeates2015}, 40 G \citep{Whitbread2017}, and 50 G \citep{Virtanen2017}.

As well as separating by magnetic field strength, it is also necessary to impose a threshold on the latitude of open field footpoints. Measurements of $B_{r, \odot}$ at the poles of the Sun are either un-obtainable due to the small tilt between the ecliptic plane and solar equator, and even when possible are challenging to measure, as only the line-of-sight (i.e. not radial) component can be directly observed. Because of these issues, the makers of synoptic maps must extrapolate values to fill in the polar regions. Different observatories use different methods, some of which result in large magnetic field values at the poles, which a simple threshold on $|B_{r}|$ would incorrectly identify as active region open field. Because (to our knowledge) no active regions have been observed at the poles of the Sun, we make the further assumption that any open field rooted at latitudes above $\pm 50^{\circ}$ is not active region open field, regardless of photospheric field strength. Two examples of this categorisation scheme are shown in the bottom two panels of Figure \ref{fig:case study}, with light colours representing coronal hole field lines and dark colours active region field lines.

\section{Results}
\label{sec:results}
\subsection{Location of Open Field Footpoints}
\begin{figure} 
	\centerline{\includegraphics[width=\textwidth]{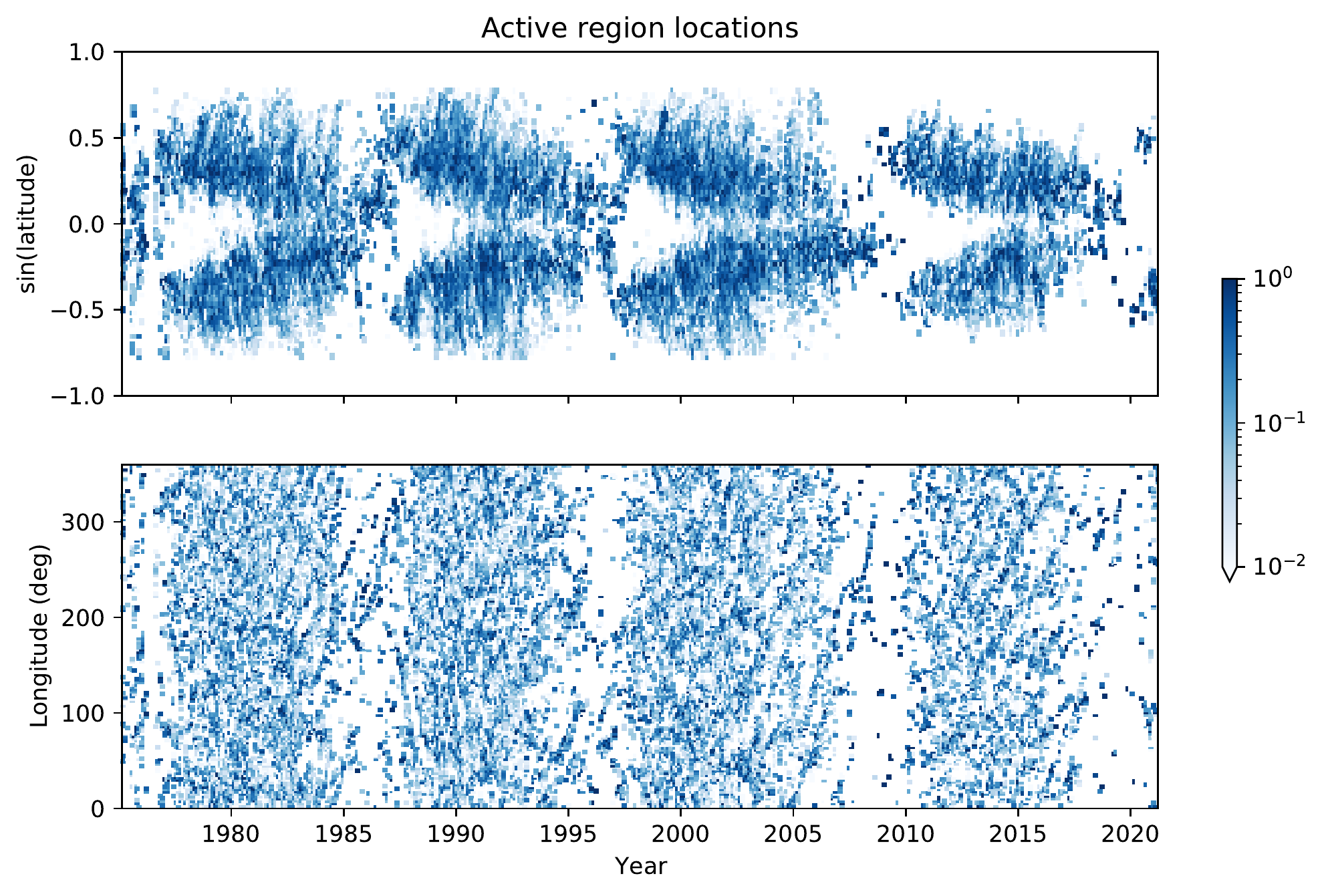}}
	\caption{Distribution of open field solar surface footpoints in active regions as a function of time, column normalised. The top panel shows distributions in sin(latitude) (summed over all longitudes), and the bottom panel distributions in longitude (summed over all latitudes). Full datasets are used from KPVT and GONG, with the gap between filled by MDI.}
	\label{fig:butterfly_ar}
\end{figure}
The latitude (top panel) and longitude (bottom panel) distributions of active region open field footpoints across four solar cycles are shown in Figure \ref{fig:butterfly_ar}. The longitude distributions reveal that AR sources are localised, typically lasting for only one or two solar rotations. A handful of regions lasted longer, and persisted for several rotations before their magnetic field dispersed to become weaker than the AR identification threshold. This agrees with observations of active region lifetimes \citep{vanDriel-Gesztelyi2015}, and we have manually checked some of the multiple-rotation trails in Figure \ref{fig:butterfly_ar} to verify that they match with active regions observable in EUV images over multiple rotations. The latitude distributions also agree with previous observations of active regions, notably the butterfly diagram that shows a reduction in active region latitudes as a solar cycle progresses \citep{Carrington1858, Maunder1922}.

\begin{figure} 
	\centerline{\includegraphics[width=\textwidth]{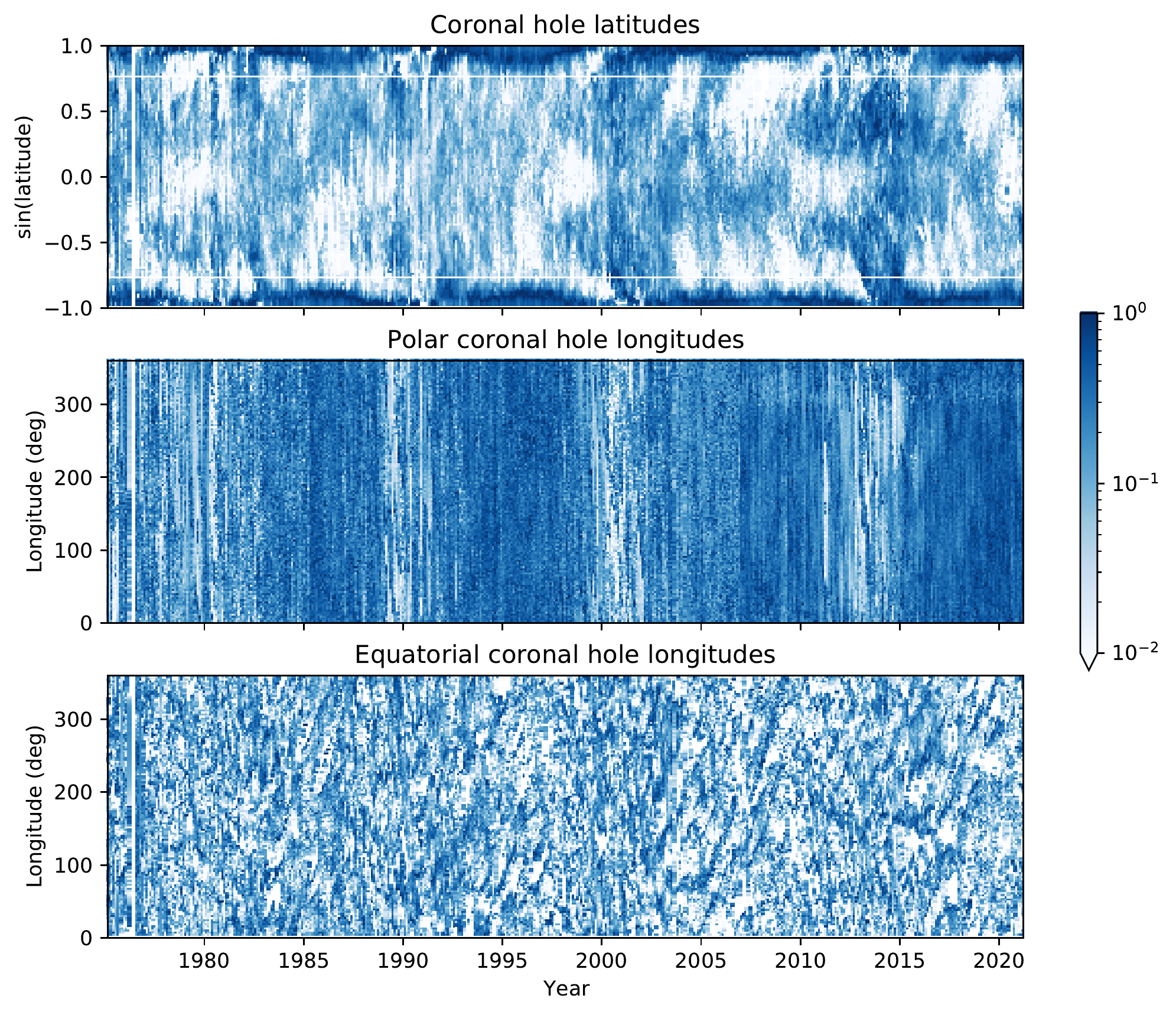}}
	\caption{Distributions of open field solar surface footpoints in coronal holes as a function of time, column normalised. The top panel shows distributions as a function of sin(latitude) (summed over longitude), with horizontal white lines at $\pm 50^{\circ}$. The middle panel shows the longitude distributions (summed over latitude) for polar coronal holes ($|\theta| > 50^{\circ}$). The bottom panel shows the longitude distributions (summed over latitude) for low latitude coronal holes ($|\theta| < 50^{\circ}$). Full datasets are used from KPVT and GONG, with the gap between filled by MDI.}
	\label{fig:butterfly_ech}
\end{figure}
The top panel of Figure \ref{fig:butterfly_ech} shows the latitude distribution of coronal hole open field footpoints. Recurring polar coronal holes are visible at high latitudes, alongside lower latitude equatorial coronal holes. Equatorial coronal holes exhibit a weak butterfly pattern, which is to be expected as they form from the diffusion of flux originally supplied by active region emergence. The middle panel of Figure \ref{fig:butterfly_ech} shows the longitude distribution of polar coronal hole open field footpoints, defined as footpoints at latitudes above $\pm 50^{\circ}$. As expected there is little longitudinal structuring of the polar coronal holes, which for the majority of the solar cycle cover the entire poles. The bottom panel of Figure \ref{fig:butterfly_ech} shows the longitude distribution of equatorial coronal holes, i.e. latitudes below $\pm 50^{\circ}$. As expected there is clear structuring in longitude, as equatorial coronal holes do not wrap all the way around the Sun. Several trails are evident in this figure, showing equatorial coronal holes that persist over multiple solar rotations. The drift of these trails in longitude with time is due to their rotation rate being slightly slower than the rotation rate used to define Carrington longitude. These trails agree with statistical results on the lifetimes of equatorial coronal holes, showing that they can exist for up to three years \citep{Hewins2020}.

Together the distributions of open field regions in active regions (Figure \ref{fig:butterfly_ar}) and coronal holes (Figure \ref{fig:butterfly_ech}), and their similarity to observational signatures of these features in EUV wavelengths give us confidence in our method for distinguishing between these two distinct solar wind source regions.

\subsection{Contribution of Active Regions to the Solar Wind}
The total open flux in each PFSS extrapolation was calculated by summing all radial magnetic field values on the source surface, and multiplying by the constant solid angle area element of each cell. The total open solar flux rooted in active regions was calculated the same way, but this time only summing over source surface pixels hosting an open field line with a photospheric footpoint rooted in an active region. The ratio of active region open flux to total open flux was taken as a proxy for their fractional contribution to all solar wind during a given solar rotation.

\begin{figure}
	\centerline{\includegraphics[width=\textwidth]{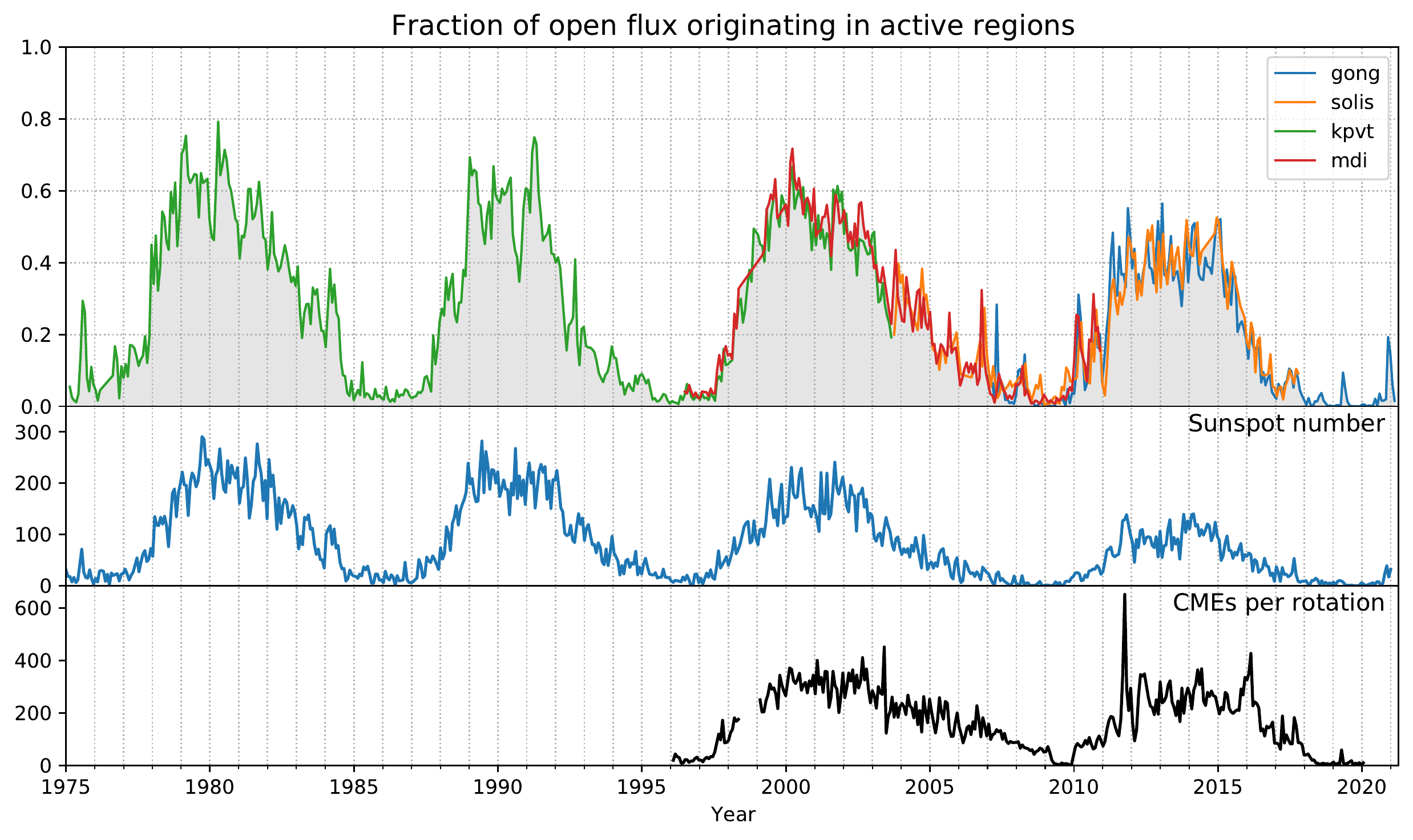}}
 	\caption{Fraction of open flux rooted in active regions as a function of time (top panel). Different observatories are denoted with different colours, as shown in the figure legend. The middle panel shows rotation-averaged sunspot number, and the bottom panel shows the number of CMEs per solar rotation taken from the LASCO CME catalogue.}
	\label{fig:all magnetograms}
\end{figure}
The fraction of total open flux rooted in active regions as a function of time is shown in the top panel of Figure \ref{fig:all magnetograms}. Results from different observatories are shown in different coloured lines. Where they overlap, the results from each observatory agree well, even on monthly timescales within individual solar cycles. The fraction of open solar flux contained within active regions follows the solar cycle; for comparison, the middle panel shows the rotation averaged sunspot number \citep{SILSOWorldDataCenter2021}. Larger amplitude sunspot cycles result in a larger fraction of open flux originating in active regions. This is a non-trivial result, depending on the details of how open flux is distributed between coronal holes (both equatorial and polar) and active regions, and how this varies with cycle amplitude. Even during the relatively weak maximum of Cycle 24, 40\% - 50\% of open flux originated in active regions, with even higher fractions during previous stronger maxima. This suggests that active regions play an equally important role to coronal holes as sources of the solar wind during solar maxima.

Because rearrangements in the global coronal magnetic field can be caused by coronal mass ejections, the bottom panel of Figure \ref{fig:all magnetograms} shows the CME rates over the last two solar cycles from the SOHO/LASCO CME catalogue \citep{Gopalswamy2009}.  CMEs do not themselves contribute significantly to the solar wind mass flux \citep{Mishra2019}, but they could play a role in opening regions of previously closed magnetic flux within active regions. Although the correlation of active region open flux with CME rate is not as strong as with sunspot number, peaks in the CME rate (e.g. in 2003, 2011, 2016) appear to show a weak association with jumps in the amount of active region open flux. In addition, variations in active region open flux between individual solar rotations are typically 10\% - 20\%. This implies that the mechanism(s) significantly modifying the active region open flux are active on timescales shorter or equal to an individual solar rotation. This is consistent with CMEs being one mechanism driving these changes in the photospheric footpoints of open magnetic flux, but more investigation is needed to show if these are causal links. 

\begin{figure}
	\centerline{\includegraphics[width=\textwidth]{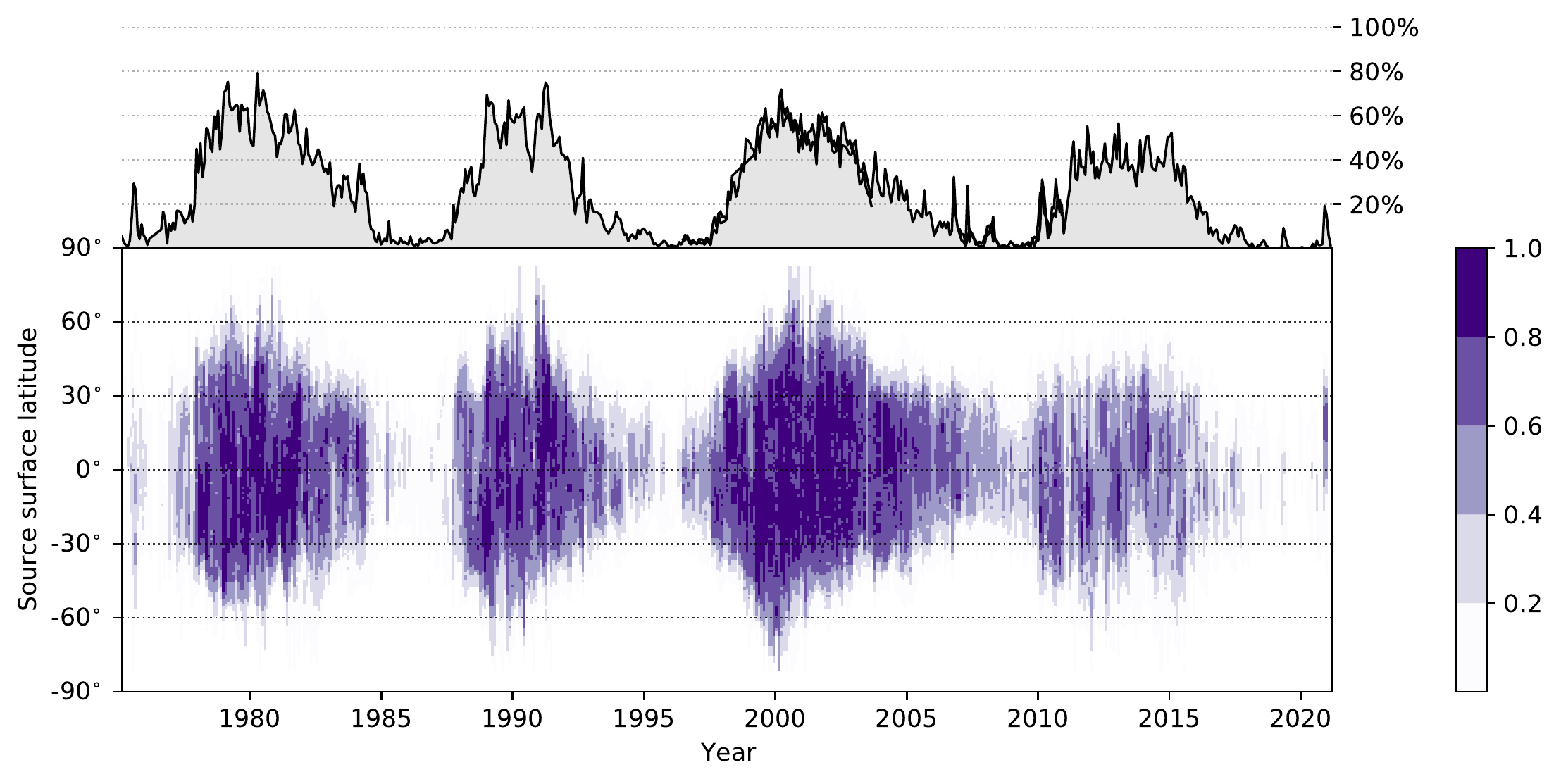}}
 	\caption{Fraction of open flux rooted in active regions as a function of source surface latitude and time, averaged over longitude. This is a proxy for solar wind latitude if solar wind propagation is radial beyond the source surface. The top panel shows the fraction averaged over all latitudes and longitudes, as previously shown in the top panel of Figure \ref{fig:all magnetograms}.}
	\label{fig:ar frac lats}
\end{figure}
As well as investigating variations over time, our dataset allows investigation of the angular extent of active region open flux in the heliosphere. Figure \ref{fig:ar frac lats} shows the fraction of all source surface longitudes connected to active region open flux, as a function of source surface latitude and time. This reveals substantial variation between different solar cycles. During Cycles 21 -- 23 active regions were connected to heliospheric latitudes $\pm 60^{\circ}$, with their extent decreasing as the solar cycle declined. In contrast, during the Cycle 24 their heliospheric connections were typically limited to $\pm 30^{\circ}$, with no obvious reduction in this extent as the cycle progressed. The gaps during solar minima were also highly variable: between both Cycles 21, 22 and the current minima there were extended $\sim$ 5 year periods with hardly any active region contributions to the solar wind, with a much shorter gap between Cycles 22, 23, and almost no gap between Cycles 23, 24.

\section{Discussion}
\label{sec:discussion}
When discussing the origins of the solar wind, it is helpful to distinguish between source regions (e.g. coronal holes, active regions) and release mechanisms (e.g. long-term open flux, interchange reconnection, closed-flux reconnection); see \cite{Viall2020} Section 2.1 for a thorough discussion of this distinction. In this paper we have used PFSS models to understand the balance of source regions between coronal holes and active regions, and how this evolves over the solar cycle (see Figure \ref{fig:all magnetograms}), but have not addressed different release mechanisms. There are several studies investigating possible release mechanisms of solar wind within individual active regions \citep[e.g.][]{Baker2007, vanDriel-Gesztelyi2012, Brooks2020}. Understanding if these mechanisms are universal properties of all active region sources combined with our new analysis of the prevalence of active region sources gives a route to understanding how solar wind release mechanisms vary over the solar cycle. In addition it should be possible to combine this information with in-situ diagnostics for solar wind origin \citep[e.g.][]{Baker2009, Stansby2019d, Owens2020} to further understand how similar or different release mechanisms are in different active regions.

A potential source of error in modelling multiple solar cycles is the necessary inclusion of magnetograms measured by a range of different observatories that are not fully inter-calibrated \citep[e.g.][]{Jones2001, Tran2005, Demidov2008, Pietarila2013, Riley2014}. We have mitigated against these differences by choosing observatory specific thresholds for identifying active region open flux (Table \ref{tab:sources}). Although some inconsistency in the fraction of active region open flux is inevitable and present, there is agreement almost always to within $\pm10\%$ between different observatories when they were observing together (Figure \ref{fig:all magnetograms}).
 
Changing the source surface radius in PFSS models can modify the exact photospheric locations of open flux, but we have shown that the fraction of open flux rooted in ARs is insensitive to this parameter, within a range of reasonable values (Figure \ref{fig:varying rss}). In contrast, varying $R_{ss}$ causes significant variation in the global distribution and location of coronal holes, meaning we could not accurately determine precise footpoints open flux to distinguish between e.g. equatorial and polar coronal holes. Further work combining our methodology with EUV observations of coronal holes \citep[e.g.][]{HessWebber2014, Hewins2020} to constrain $R_{ss}$ could be used to remove this limitation.

We have hypothesised that the mechanism opening up new active region open flux is interchange reconnection in the corona. There is evidence that this coronal process can occur gradually and continuously \citep[e.g.][]{Higginson2017}, or as as more discrete CME events \citep[e.g.][]{Cohen2009, Driel-Gesztelyi2014}, but it is not clear which of these processes dominates on a global scale. It is not possible to directly investigate this question within PFSS models, which are time-independent snapshots of the coronal field. Time evolving magnetogfrictional simulations could be used however, where the boundary photospheric magnetic field is evolved in time to drive coronal magnetic field evolution \citep{Yeates2014}. These models have previously been used to predict where flux ropes form and erupt in the corona \citep{Yeates2014, Lowder2017}, but it should also be possible to identify interchange reconnection occurring in these simulations, giving a possible avenue into investigating the mechanisms behind the opening and closing of active region open flux.

The fraction of solar wind originating in active regions could have implications for the properties of the solar wind, and how this affects the heliosphere across the solar cycle. Backmapping solar wind measured at 1 AU to active region sources identified in EUV shows that active region solar wind is slower than the average solar wind \citep{Fu2015, Zhao2017}. Active regions are hotter than coronal holes, which drives increased mass fluxes in the corona \citep{Stansby2020f}, but in the case of the Sun the magnetic field expansion almost exactly cancels this difference out, resulting in a remarkably constant solar wind mass flux \citep{Wang2010} that is independent of source type. There is still plenty of scope for further investigation of the properties of solar wind originating in active regions. When coupled with our results on the time-latitude distribution of active region sources (Figure \ref{fig:ar frac lats}) this could provide a way to predict heliospheric conditions where active region solar wind dominates.

\section{Conclusions}
\label{sec:conclusions}

Using global potential field source surface modelling of the corona (Section \ref{sec:methods}), we have estimated what fraction of solar wind originates in active regions, as a function of latitude and time. The fractional contribution of active regions to the solar wind is negligible at solar minimum, and typically 40\% - 60\% at solar maximum, scaling with sunspot number (Figure \ref{fig:all magnetograms}). The latitudinal extent of active region solar wind is highly variable between different solar cycles (Figure \ref{fig:ar frac lats}): in Cycles 21 - 23 active region wind extended to $\pm 60^{\circ}$, but during the weaker Cycle 24 typically only reached $\pm 30^{\circ}$. Even if the upcoming Cycle 25 is a weak cycle, Parker Solar Probe \citep{Fox2016} and Solar Orbiter \citep{Muller2020} will observe an increasing amount of solar wind from active region sources, allowing further understanding of the properties and release mechanisms of active region solar wind.

\appendix   
\section{Varying Source Surface Radius}
\label{app:varying rss}
\begin{figure}
	\centerline{\includegraphics[width=\textwidth]{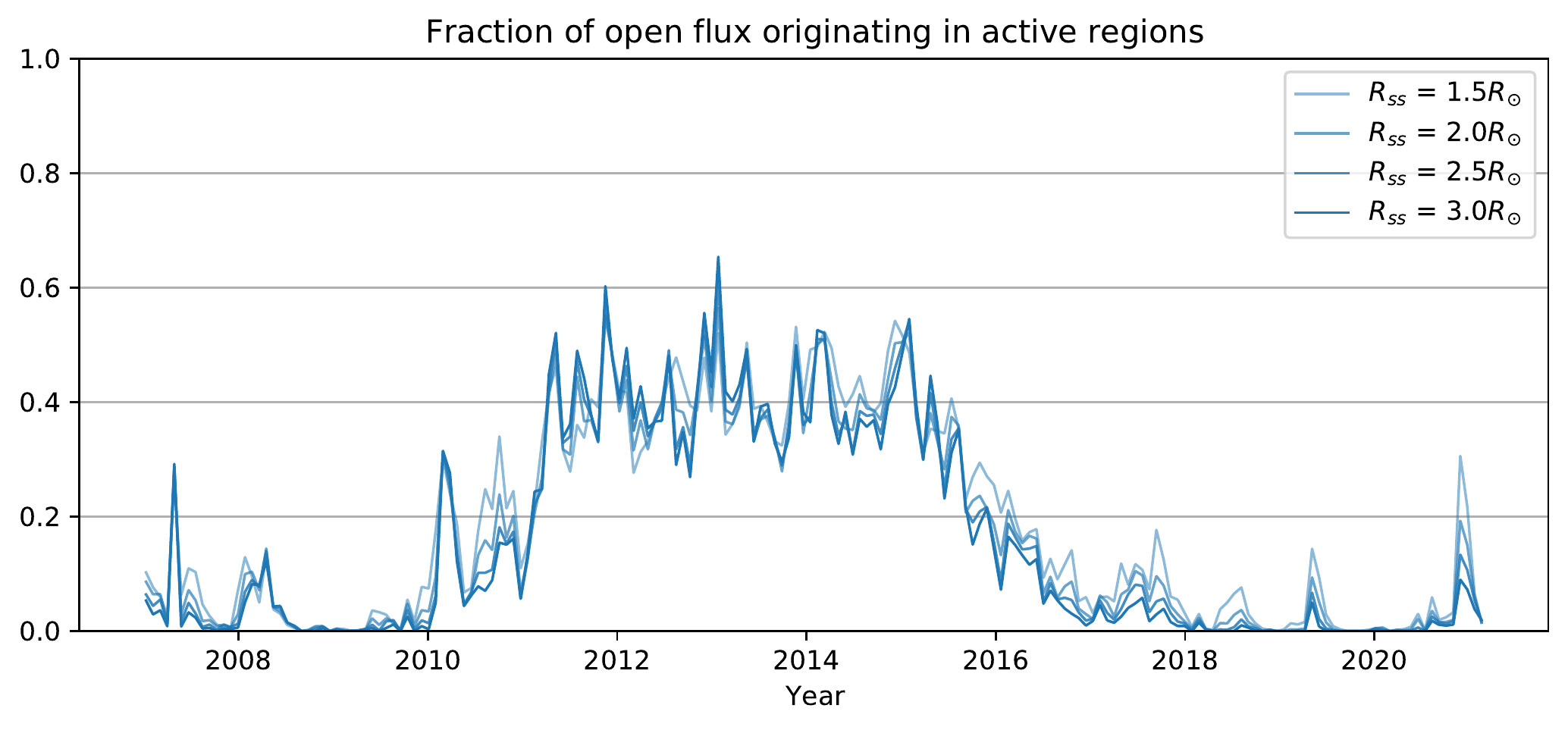}}
	\caption{The effect of  varying $R_{ss}$ on estimates of the fraction of open flux rooted in active regions. All data are from GONG, with different $R_{ss}$ values indicated by different line weights from light (low $R_{ss}$) to dark (high $R_{ss}$).}
	\label{fig:varying rss}
\end{figure}

To check that the results in Figure \ref{fig:all magnetograms} are not affected by model parameters, Figure \ref{fig:varying rss} shows the fraction of open flux originating in active regions for PFSS model source surface heights ranging from 1.5$R_{\odot}$ (thinnest line) to 3.0$R_{\odot}$ (thickest line). While for some rotations there is a slight tendency for smaller source surface radii to result in a slightly larger fraction of flux rooted in active regions, this is a small effect and does not change any of the conclusions in this paper.

\begin{acks}
D.S. is supported by STFC grant ST/S000240/1. L.v.D.G. is partially funded under under STFC consolidated grant number ST/S000240/1 and acknowledges the Hungarian National Research, Development and Innovation Office grant OTKA K-131508. L. M. G. is grateful to the Royal Society for support through a University Research Fellowship. T. S. H. is supported by STFC grant ST/S000364/1. The authors thank the referee for constructive comments that improved the paper. This research used the following software packages: pfsspy \citep{Stansby2020e}, sunpy \citep{TheSunPyCommunity2020}, astropy \citep{TheAstropyCollaboration2018}, Matplotlib \citep{Hunter2007}.  This work utilises SOLIS data obtained by the NSO Integrated Synoptic Program (NISP), managed by the National Solar Observatory, which is operated by the Association of Universities for Research in Astronomy (AURA), Inc. under a cooperative agreement with the National Science Foundation. NSO/Kitt Peak data used here are produced cooperatively by NSF/NOAO, NASA/GSFC, and NOAA/SEL. This work utilises data from the National Solar Observatory Integrated Synoptic Program, which is operated by the Association of Universities for Research in Astronomy, under a cooperative agreement with the National Science Foundation and with additional financial support from the National Oceanic and Atmospheric Administration, the National Aeronautics and Space Administration, and the United States Air Force. The GONG network of instruments is hosted by the Big Bear Solar Observatory, High Altitude Observatory, Learmonth Solar Observatory, Udaipur Solar Observatory, Instituto de Astrof\'{\i}sica de Canarias, and Cerro Tololo Interamerican Observatory. SOHO is a project of international cooperation between ESA and NASA. This CME catalog is generated and maintained at the CDAW Data Center by NASA and The Catholic University of America in cooperation with the Naval Research Laboratory.
\end{acks}

\section*{Ethics declarations}
\subsection*{Declaration of Potential Conflicts of Interest}
The authors declare that they have no conflicts of interest.

\bibliographystyle{spr-mp-sola}
\bibliography{/Users/dstansby/Dropbox/zotero_library}  

\end{article} 
\end{document}